\documentclass[preprint]{aastex}
\usepackage{graphicx}
\usepackage{amsmath}

\begin{document}

\title{Star Formation Properties in Barred Galaxies(SFB). III. Statistical Study of Bar-driven Secular Evolution using a sample of nearby barred spirals}
\author{Zhi-Min Zhou\altaffilmark{1}, Chen Cao\altaffilmark{2,3,4}, and Hong Wu\altaffilmark{1}}
\altaffiltext{1}{Key Laboratory of Optical Astronomy, National Astronomical Observatories, Chinese Academy of Sciences, Beijing 100012, China; zmzhou@bao.ac.cn, hwu@bao.ac.cn}
\altaffiltext{2}{School of Space Science and Physics, Shandong University at Weihai, Weihai, Shandong 264209, China; caochen@sdu.edu.cn.}
\altaffiltext{3}{Shandong Provincial Key Laboratory of Optical Astronomy \& Solar-Terrestrial Environment, Weihai, Shandong 264209, China}
\altaffiltext{4}{Visiting Scholar, Infrared Processing and Analysis Center (IPAC), California Institute of Technology, Pasadena, CA 91125}
\shortauthors{Zhou, Cao, \& Wu}
\shorttitle{Statistical Study of Barred Galaxies}

\begin{abstract}
Stellar bars are important internal drivers of secular evolution in disk galaxies. Using a sample of nearby spiral galaxies with weak and strong bars, we explore the relationships between the star formation feature and stellar bars in galaxies. We find that galaxies with weak bars tend to be coincide with low concentrical star formation activity, while those with strong bars show a large scatter in the distribution of star formation activity. We find enhanced star formation activity in bulges towards stronger bars, although not predominantly, consistent with previous studies. Our results suggest that different stages of the secular process and many other factors may contribute to the complexity of the secular evolution. In addition, barred galaxies with intense star formation in bars tend to have active star formation in their bulges and disks, and bulges have higher star formation densities than bars and disks, indicating the evolutionary effects of bars. We then derived a possible criterion to quantify the different stages of bar-driven physical process, while future work is needed because of the uncertainties.

\end{abstract}

\keywords{galaxies: general --- galaxies: photometry --- galaxies: evolution --- galaxies: spiral --- galaxies: structure}

\section{Introduction}
Stellar bars are ubiquitous in disc galaxies. Observed bar fractions range from one to two thirds in the local Universe, depending on the method to identify bars and the sample and wave-bands observed \citep{Eskridge00,Menendez07}. Bars are also found in galaxies up to z $\sim$ 1 \citep{Sheth08,Melvin14}.
Bars are one of the major internal drivers of secular evolution in galaxies \citep[see][for review]{KK04}. The nonaxisymmetric gravitational potentials of bars can redistribute the angular momentum of gas and stars in galactic disks, induce large-scale streaming motions \citep{Regan95,Regan99,Sakamoto99b, Sheth05}, and result in an increase in gas mass and enhancement of star formation activity in the galactic central regions \citep{Sersic65,Hawarden86,Ho97}.

Many works, both theoretical and observational, have been made to explore the process of Bar-driven secular evolution. Numerical simulations have established that the gravitational torque of large-scale bars can make gas lose angular momentum, and allow it to inwards towards the galactic center \citep{Athanassoula92, Sellwood93, Piner95, Athanassoula03}. Observations indicate that barred spirals have higher molecular gas concentrations in their central kiloparsec than unbarred systems \citep{Sakamoto99b, Sheth05}. The younger stellar populations and higher star formation rates (SFRs) in the bulges of barred galaxies are also found when compared with unbarred ones \citep[e.g.,][]{Sersic65,Huang96, Alonso01,Kormendy05,Fisher06,Coelho11,Wang12}.
\citet{KK04} showed the correlations between the surface densities of SFR and gas, and compared these correlations in circumnuclear star-forming rings in barred galaxies with those in disks of spiral galaxies. They found that both SFR densities and gas densities in circumnuclear rings are higher than those in the outer disks. 

However, some observations found conflicting results as to how bar affected the evolution of disk galaxies. \citet{Ho97} showed that the bars in late-type spirals have no noticeable impact on the likelihood of a galaxy for hosting nuclear star formation. Furthermore, there are some barred galaxies with no molecular gas detected in their nuclear regions where it may have been consumed by star formation\citep{Sheth05}. Using emission-line diagnostics to identify central star formation activity in galaxies, \citet{OH11} found that bar effects on central star formation seem higher in redder galaxies and are more pronounced with increasing bar length, but the effects on central star formation are not visible in blue galaxies. Similarly, \citet{Wang12} found only strong bars result in enhanced central star formation in galaxies, but not weak bars do. \citet{Ellison11} found enhanced fibre SFR and metallicities just exist in massive barred galaxies, and no significant correlation between SFR enhancement and bar length or ellipticity. Based on the Galaxy Zoo 2 data set, \citet{Cheung13} studied the behavior of bars in disk galaxies and found that the trends of bar likelihood and bar length with bulge prominence are bimodal with specific star formation rates (SSFRs). 

In previous studies, comparisons were often made between barred and unbarred galaxies, whereas the comparisons between the variation in stellar bar properties and their associated star formation are crucial to the understanding of the bar-driven secular process in galaxies. Thus, large sample and multi-wavelength data are required. In our previous work, three barred galaxies in different evolutionary stages were analyzed based on the data in different wavelength bands, they are NGC 7479 in \citet[][, hereafter Paper I]{Zhou11}, NGC 2903 and NGC 7080 in \citet[][, hereafter Paper II]{Zhou12}. In this paper, we use multi-wavelength data from optical to infrared (IR) to analyzes a relatively large sample of nearby barred galaxies (with both weak and strong bars, including from early to late Hubble types). We aim to statistically study the correlation between star formation activity and bar properties, especially its strength. 

This paper is organized as follows. In Section 2, we describe the selection of the sample and give a brief introduction of the data. In Section 3, we present the method used for image decomposition, and derive the structural characterization of our galaxies. We explore the correlations between the star formation and bar properties in Section 4. Then we discuss our results and explore the possible evolutionary episodes of bar-driven secular evolution in Section 5. Finally, we finish with our summary and conclusions in Section 6.

\section{Sample and Data}
\subsection{Sample Selection}
Our sample is mainly selected from three {\it Spitzer} local programs: the Spitzer Infrared Nearby Galaxies Survey \citep[SINGS;][]{Kennicutt03}, the Local Volume Legacy \citep[LVL;][]{Lee08, Dale09}, and the Infrared Hubble Atlas \citep{Pahre04}. From all nearby galaxies in these programs, we selecte galaxies with bars, i.e., with Hubble type of SB and SAB. In addition, we apply an inclination cut to remove apparently edge-on galaxies (inclination $>$ 65$^\circ$ or ellipticity $<$ 0.58), as well as removing objects that are obviously merging. This produce the final sample size of 50 galaxies.

\begin{figure}[!htop]
\center
\includegraphics[angle=0,width=\textwidth]{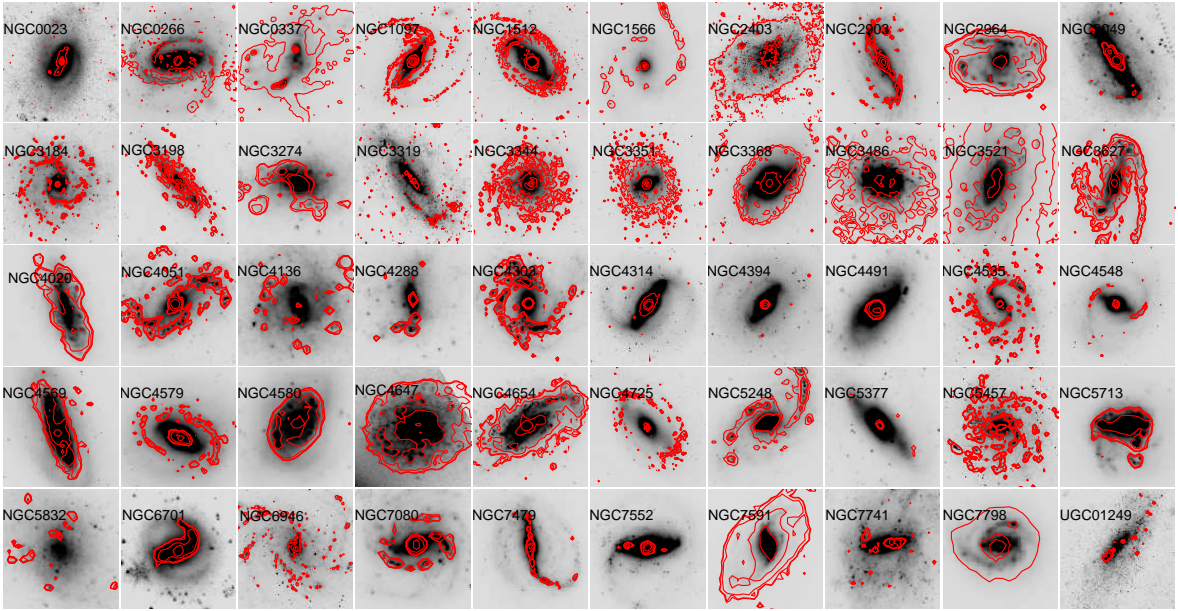}
\caption{The mass images of the galaxies in our sample. Overplotted are the SFR imaging contours, which are in arbitrary units, optimized to show and identify the structures of the star formation complexes. The galaxy names are specified in the panels, and north is up and east to the left. 
\label{fig1}}
\end{figure}

\begin{figure}[!htop]
\center
\includegraphics[angle=0,width=0.45\textwidth]{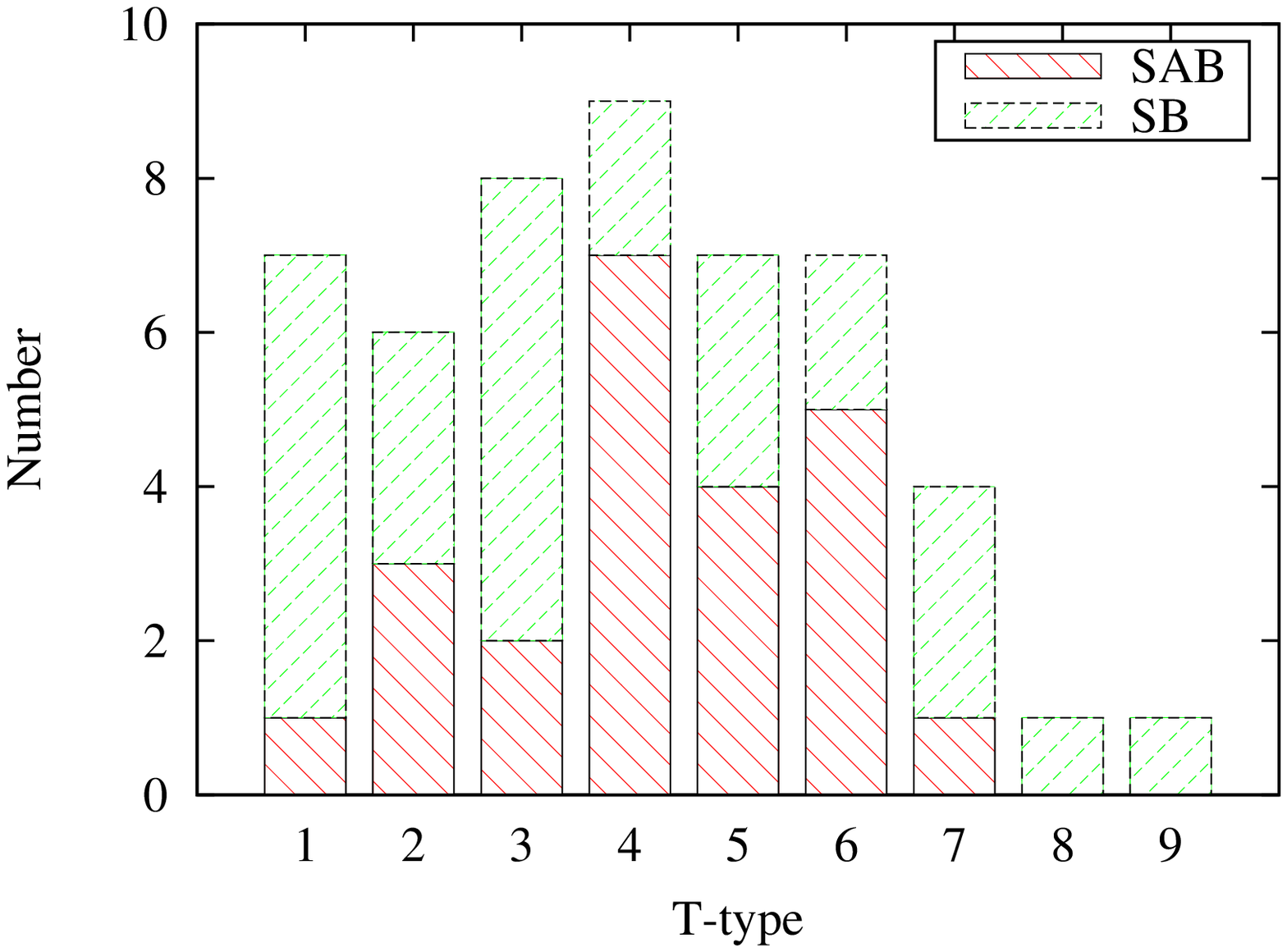}
\includegraphics[angle=0,width=0.45\textwidth]{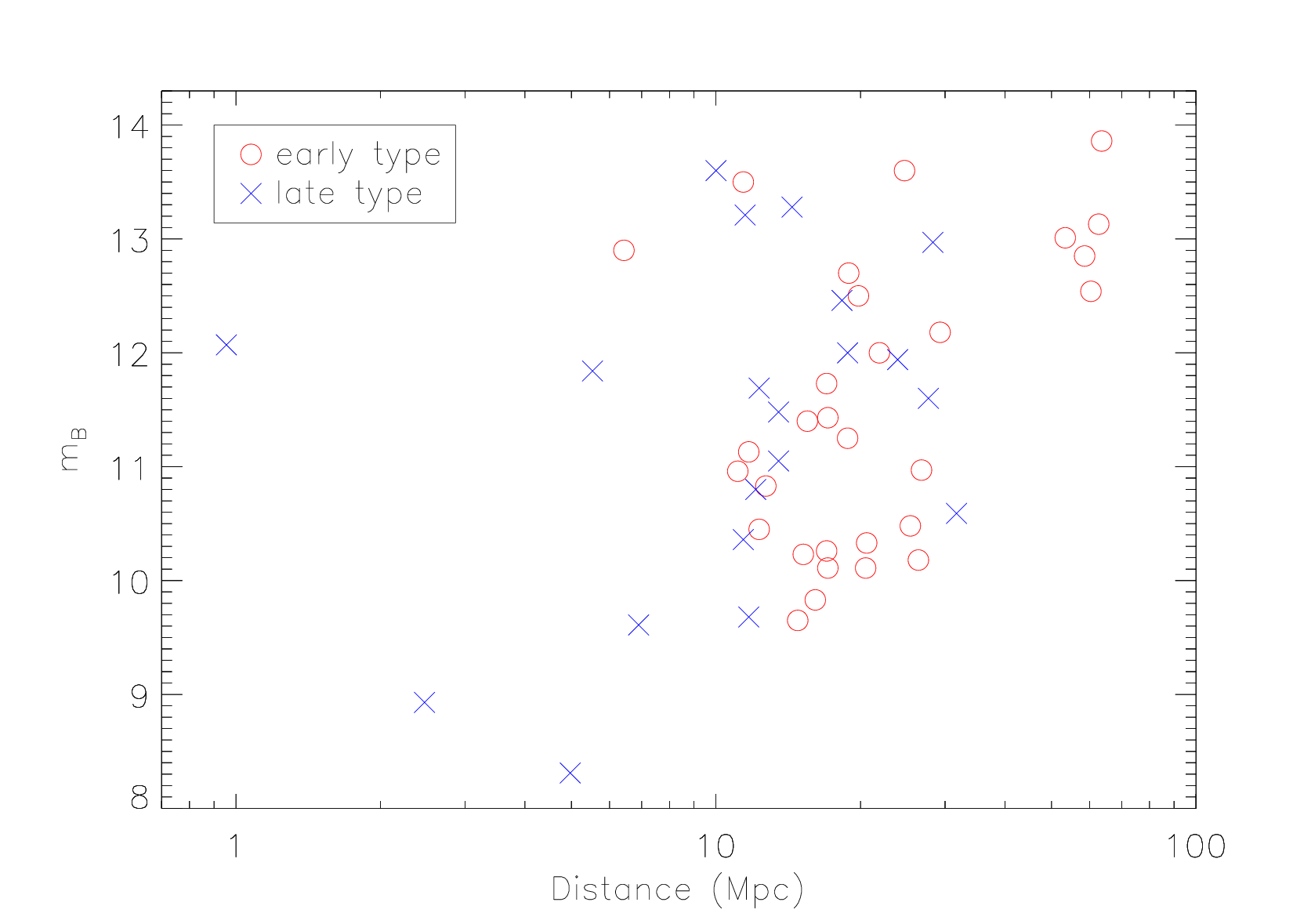}
\caption{Morphological distribution ({\it left}) and magnitude-distance distribution ({\it right}) of our sample. In the {\it left} panel, the morphological parameter T = 1--9 corresponds to the Hubble types Sa--Sm, and the sample is divided into SB and SAB galaxies. In the {\it right} panel, the distance used is luminosity distance from NED, and the sample is divided into early (T = 1--4) and late (T = 5--9) types, red circles for early-type galaxies and blue crosses for late-type ones.
\label{fig2}}
\end{figure}

Table~\ref{table1} shows the general properties of the 50 galaxies in our sample (see Figure~\ref{fig1}). The distances given in Table~\ref{table1} were luminosity distances from the NASA Extragalactic Database (NED\footnote{http://nedwww.ipac.caltech.edu/}). The distance of the studied sample has mean value of 20.5 Mpc, and the maximum value of 63.6 Mpc. The standard isophotal galaxy diameters (D$_{25}$) are $\sim$ 3$'$, with all sizes larger than 1$'$. Galaxies in the sample include a wide variety of Hubble types.
There are 27 SB galaxies and 23 SAB galaxies ranging from Sa to Sm, corresponding to the numerical morphological index T = 1--9. Figure~\ref{fig2} shows the distribution of morphology for all sample galaxies. In the following analysis we divided galaxies into early-type (T $\leq$ 4) and late-type (T $\geq$ 5) spirals. Thus, there are 30 early-type spirals and 20 late-type spirals in our sample. In the right panel of Figure~\ref{fig2}, we also plotted the distance and magnitude distribution of our sample.

\subsection{Data Acquisition and Reduction}
In order to investigate the properties of barred galaxies, we archived infrared images from the {\it Spitzer Space Telescope} \citep{Werner04} data archive. In optical wavelength, the narrow-band H$\alpha$ images from ground-based telescopes are collected and used to analyze the star formation activity in galaxies. In the following section, we provide the multi-wavelength data observation/acquisition and a brief description of data reduction. A more detailed description of data acquisition and reduction can be found in Paper I and II.


\subsubsection{Optical Data}
In our sample, 17 galaxies were observed with the 2.16m telescope at Xinglong Observatory of the National Astronomical Observatories of the Chinese Academy of Sciences\footnote{http://www.xinglong-naoc.org/English/216.html} during 2009--2010. The instrument we used was the BAO Faint Object Spectrograph and Camera (BFOSC), it had a Lick 2048$\times$2048 CCD detector with a plate scale of 0\farcs305 pixel$^{-1}$ and a field of view of $\sim$ 10$'$ $\times$ 10$'$. The galaxies were imaged with broadband R filter and narrowband filters centered on the appropriately redshift H$\alpha$+[N {\sc ii}] $\lambda\lambda$6563, 6583 emission lines, the former images were used to subtract stellar continuum from the H$\alpha$+[N {\sc ii}] images. The exposure time was typically 600s for R-band and 3000$-$3600s for H$\alpha$-band. The detailed information about the observation is listed in Table~\ref{table2}.

The IRAF\footnote{IRAF is the Image Reduction and Analysis Facility written and supported by the IRAF programming group at the National Optical Astronomy Observatories (NOAO) in Tucson, Arizona which is operated by AURA, Inc. under cooperative agreement with the National Science Foundation} package was used for data reduction. The reduction pipeline includes checking images, adding keywords to fits headers, subtracting overscan and bias, correcting bad pixels and flat-fielding, along with removing cosmic-rays. The flux calibration of broadband images was made using the standard stars selected from the Landolt fields \citep{Landolt92} and observed with the corresponding filter on the same night. After sky subtraction, the scaled R-band images were subtracted from the aligned narrowband images to obtain continuum-free H$\alpha$+[N {\sc ii}] line images, the scale factors were calculated using photometric count ratios of several ($\sim$ 5) unsaturated field stars in both filter images. The narrowband images were flux calibrated based on the scale factors of R-band images and effective transmissions of narrowband and R-band filters. Comparing the calibrated H$\alpha$ and R-band images indicates that there is $\sim$ 6\% of H$\alpha$ emission lost in the process of removing stellar continuum, while the contribution from [N {\sc ii}] to the total H$\alpha$ flux is typically of order 10\% or less \citep{Kennicutt08}. Thus, the final images have spatial resolutions of $\sim$ 2$''$ along with flux uncertainty of $\sim$ 10\%.

In addition to our observations, optical data in R broadband and H$\alpha$ narrowband filters were gathered in other previous studies, which are listed in Table~\ref{table3}. Images of 16 galaxies from SINGS sample were obtained from the ancillary data of the legacy project, which were observed by the Kitt Peak National Observatory (KPNO) 2.1 m telescope and the Cerro Tololo Inter-American Observatory (CTIO) 1.5 m telescope. Images of 8 galaxies from LVL sample were obtained from the ancillary data of LVL project, as part of the precursor 11 Mpc H$\alpha$ Survey \citep{Kennicutt08}. The optical data of other sample galaxies were derived from NED. The observations, data processing, flux calibration of the optical archival images were nicely matched the counterparts of our own data.

\subsubsection{IR Data}
To investigate the infrared properties of our sample, we retrieved the available archived {\it Spitzer} near/mid-IR data from SINGS and LVL Program Datasets, including the Infrared Array Camera \citep[IRAC;][]{Fazio04} imaging data at 3.6, 4.5, 5.8, and 8.0 $\mu$m bands, as well as the Multiband Imager and Photometer for Spitzer \citep[MIPS;][]{Rieke04} imaging data at 24 $\mu$m band. The images of IRAC each channel had a pixel scale of 0\farcs75, and the spatial resolutions of $\sim$ 2\farcs0, which are similar to those of optical H$\alpha$ images. The MIPS at 24 $\mu$m images had a pixel scale of 1\farcs5, with the full width at half-maximum (FWHM) of $\sim$ 6\farcs0.

For our galaxies not from SINGS or LVL surveys, we obtained the Basic Calibrated Data (BCD) generated from the {\it Spitzer} data reduction pipeline, and then mosaicked them with the MOsaicker and Point source Extractor (MOPEX) software. The final images were produced with the same plate scales and similar spatial resolutions to the archived infrared data.

In images of IRAC 8.0 $\mu$m band, the flux is mainly from the polycyclic aromatic hydrocarbon (PAH) emission, dust-continuum emission \citep{Pahre04,Pahre04b}
But there is the contribution of stellar continuum in this band. To remove the stellar contribution, the scaled IRAC 3.6 $\mu$m band images were used as the stellar continuum with the scale factor of 0.232 \citep{Helou04}, and were subtracted from 8.0 $\mu$m images. Hereafter we referred the continuum-free 8 $\mu$m dust emission as to 8$\mu$m(dust).

\section{Image Decomposition and Bar characterization}
\label{sec3}
We derive the structural parameters of bars, i.e., the length, ellipticity and strength, with structural decomposition of barred galaxies. The IRAC 3.6 $\mu$m images, whose the light is mainly coming from M and K giants \citep{Sheth10}, can nicely trace stellar mass distribution in nearby galaxies, although there is contamination from the 3.3 $\mu$m PAH, hot dust and intermediate-age stars \citep{Meidt12}. Therefore, we used the image-fitting code BUDDA v2.2 \citep{de04, Gadotti08} to perform a two-dimensional (2-D) decomposition of 3.6 $\mu$m images in bar/bulge/disc stellar components. The present version of BUDDA allows us to decompose a galaxy into an exponential disk, a bulge with a S\'ersic profile, and a bar \citep[e.g.,][]{Kim14}. In the decomposition results, we obtain the galaxy structural parameters, such as the size, ellipticity and position angle (PA) of each structure, which were also used to define the regions of the bulge, bar and disk in one galaxy.

\begin{figure}[!htop]
\center
\includegraphics[angle=0,width=0.8\textwidth]{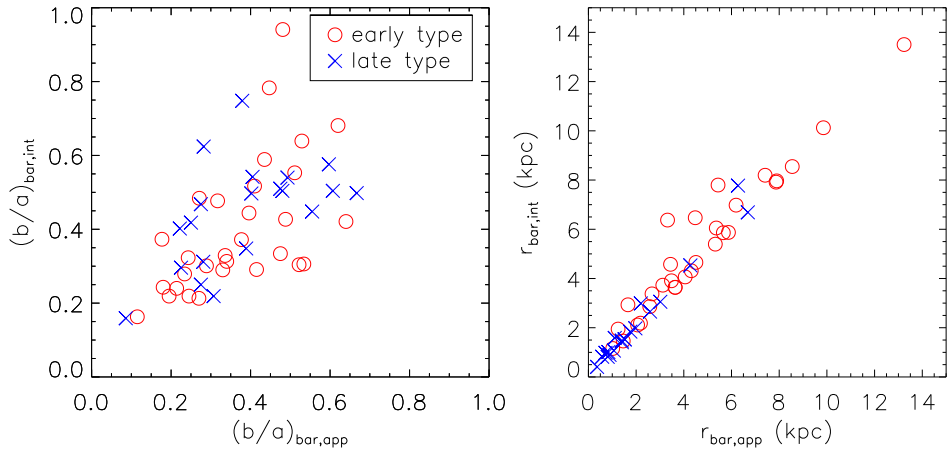}
\caption{Comparison of bar properties between intrinsic and apparent values. The intrinsic axial ratio (b/a)$_{bar,int}$ ({\it left}) and length r$_{bar,int}$ ({\it right}) of bars were calculated with the galaxy inclination and angles between the axes and line of nodes. 
\label{fig3}}
\end{figure}

The structural parameters we obtain may be affected by inclination effects. To derive the intrinsic physical parameters, we make the corrections for the apparent values following the method of \citet{Martin95}. In \citet{Martin95}, the galaxy inclination {\it i} was defined with the axial ratio of the isophote 25 mag/arcsec$^2$ in B-band obtained from the Third Reference Catalogue of Bright Galaxies \citep[RC3;][]{de91}, and the assumption that the galaxy is intrinsically axisymmetric at large radii. Given that the bars are thin 2-D structures, the intrinsic axial ratio (b/a)$_{bar}$ and length r$_{bar}$ (the semi-major axis) of bars were calculated with the galaxy inclination and angles between the axes and line of nodes. 
Figure~\ref{fig3} compares the bar properties before and after correction. There is large correction on the axial ratio of bars, while there is excellent consistency between the length of bars.

\begin{figure}[!htop]
\center
\includegraphics[angle=0,width=\textwidth]{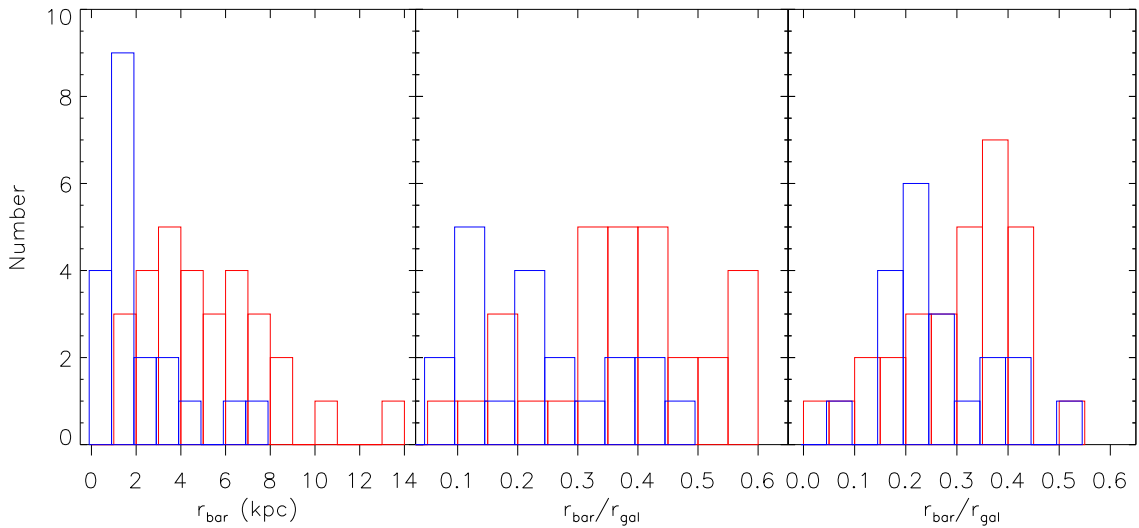}
\caption{Distributions of bar properties of early- (red line) and late- (blue line) type spirals in our sample. The bar length ({\it left}) is indicated by the physical length of the bar semi-major axis, the normalized value ({\it middle}) is indicated by the ratio between the physical length of the bar semi-major axis and the galaxy radii at 25 mag/arcsec$^2$, and the bar strength ({\it right}) is indicated by the ellipticity-based $f_{bar}$ parameter defined by \citet{Abraham00}. In each panel, two histograms are slightly offset in x-axis for purposes of clarity.
\label{fig4}}
\end{figure}

Figure~\ref{fig4} shows the distribution of the intrinsic bar lengths of the sample, both absolute and normalized. The latter is relative to the galaxy size R$_{25}$, which is measured by the semi-major axis of the isophote at the brightness of 25 mag/arcsec$^2$ in B-band from RC3. The mean length of stellar bars is 5.2 kpc and 0.37 R$_{25}$ in early types, 2.2 kpc and 0.23 R$_{25}$ in late types, consistents with earlier studies \citep{Erwin05, Menendez07}.

\begin{figure}[!htop]
\center
\includegraphics[angle=0,width=0.8\textwidth]{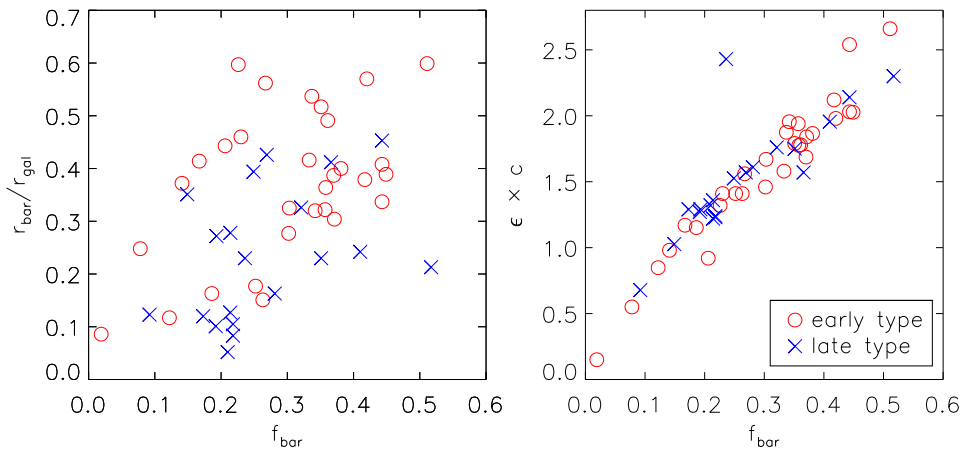}
\caption{Comparisons between the bar strength parameter $f_{bar}$ and other two indicators of bar strength: the normalized bar size $r_{bar}/r_{gal}$ ({\it left}) and combination of bar ellipticity and boxiness $\epsilon \times {\it c}$ ({\it right}). The symbols are the same as the right panel of Figure~\ref{fig2}. 
\label{fig5}}
\end{figure}

Bar strength represents the ability of driving gas into galactic center, and is usually estimated by bar structural parameters, such as the ellipticity and length of the bar \citep[e.g.,][]{Jogee04, Laurikainen07}. A more direct parameter is the gravitational torque Q$_g$, which is the maximum relative tangential force in the bar region, normalized to the underlying mean axisymmetric force field \citep{Buta01}. Here, another bar strength parameter is used, it is defined following \citet{Abraham00}:
 \begin{equation}
 f_{bar}=\frac{2}{\pi}[arctan(b/a)_{bar}^{-1/2}-arctan(b/a)_{bar}^{+1/2}],
 \label{eq1}
 \end{equation}
 where (b/a)$_{bar}$ is the intrinsic axial ratio of the bar. The ellipticity-based parameter $f_{bar}$ has been used in many previous studies \citep[e.g.,][]{Whyte02, Aguerri09}, and is proved to be well correlated with Q$_g$ \citep{Buta04, Laurikainen07}. The right panel of Figure~\ref{fig4} shows the distribution of the bar strength $f_{bar}$. Bars in early-type galaxies are clearly stronger then those in late-type galaxies. Besides the parameter $f_{bar}$ and  bar torque, the normalized bar size $r_{bar}/r_{gal}$ and the combination of bar ellipticity and boxiness $\epsilon \times {\it c}$ were also used in \citet{Gadotti11} to estimate bar strength. We compare the two parameters with $f_{bar}$ in Figure~\ref{fig5}. This figure shows that $f_{bar}$ is well consistent with the other two parameters, especially $\epsilon \times {\it c}$, so it is used as the estimation of bar strength in the following analysis. Their values are also listed in Table~\ref{table4}.

\section{Relation between Galactic Star Formation Properties and Bars}
Star formation activity is important feature in our galaxies. In our data of the sample, the H$\alpha$ emission from ground-based observations and the 8 $\mu$m(dust) and 24 $\mu$m emission from {\it Spitzer} observations can be used as star formation tracers \citep{Kennicutt98, Wu05, Calzetti07, Zhu08, Kennicutt09, Hao11}. To derive extinction-corrected SFR distribution, we used the combination of H$\alpha$ and 8 $\mu$m(dust) emissions: the former measuring the ionization rate, and the latter tracing the dust-obscured component of the star formation. The calibration for the composite indicators is followed the form in \citet{Zhu08}:
\begin{equation}
SFR_{H\alpha +8\mu m}(M_{\odot}~yr^{-1})=7.9\times 10^{-42}[L(H\alpha)_{obs}+0.010L(8)](erg~s^{-1}) \label{eq_SFR},
\end{equation}
which adopts the \citet{Salpeter55} initial mass function (IMF) with a slope −2.35 for stellar masses in the range 0.1--100 M$_{\odot}$. 

In addition to the star formation activity, we also paid attention to the stellar mass of galaxies to derive the stellar properties. Here, the IRAC 3.6 $\mu$m emission is used as a proxy of the stellar mass, since its light is dominated by the photospheric emission from low-mass stars as mentioned above \citep{Smith07, Hancock07, Wu07, Li07}. The calibration is from \citet{Zhu10} (their Eq. [2]):
\begin{equation}
Log_{10} {~M(M_{\odot})}=(-0.79\pm0.03)+(1.19\pm0.01)\times Log_{10} { ~\nu L_{\nu}{[3.6\mu m]}(L_{\odot})}\label{eq_mass},
\end{equation}
which has proved to be a good stellar mass tracer compared with K$_s$-band luminosity. Using these equations, we explored the correlation between the star formation activity and stellar bars in our galaxies. 

\subsection{Star Formation and Stellar Mass Distribution}
\label{sec4.1}
In order to describe the global properties of the galaxies, we first obtained the mass images and SFR images by combining the relevant images based on Equation~\ref{eq_mass} and \ref{eq_SFR}, respectively. All images are listed in Figure~\ref{fig1}, along with the distributions of stellar mass and star formation shown. Then we calculated the rotational asymmetry and concentration index of the mass and SFR images to quantify their distributions. The asymmetry describes how symmetric a galaxy system is, and the concentration parameter measures how compact the light distribution is. Both indexes are used to describe galaxy morphology in many studies \citep[e.g.,][]{Bershady00, Conselice00, Abraham03, Gil07, Munoz09}. The former is calculated by comparing the original image of a galaxy with its rotated counterpart (the usual rotation angle is 180$\degr$). Here, we followed the method in \citet{Conselice00}:
\begin{equation}
A=\frac{\sum{\left| I_{180\degr}-I_0 \right|}}{2\sum{\left| I_0 \right|}},
\end{equation}
where I$_0$ is the intensity distribution in the original image and I$_{180\degr}$ is the intensity distribution in the rotated image with rotation angle of 180$\degr$.

The concentration index is usually calculated by the ratio of two radii containing fixed fractions of the total flux in a galaxy. We adopt the index C$_{42}$ defined by \citet{Kent85}:
\begin{equation}
C_{42}=5log_{10}(r_{80}/r_{20}),
\end{equation}
where r$_{80}$ and r$_{20}$ are the semi-major axes of the ellipses containing 80\% and 20\% of the total luminosity, respectively. The resulting structural indexes for the distributions of star formation and stellar mass in galaxies are listed in Table~\ref{table4}. 

\begin{figure}[!htbp]
\center
\includegraphics[angle=0,width=0.8\textwidth]{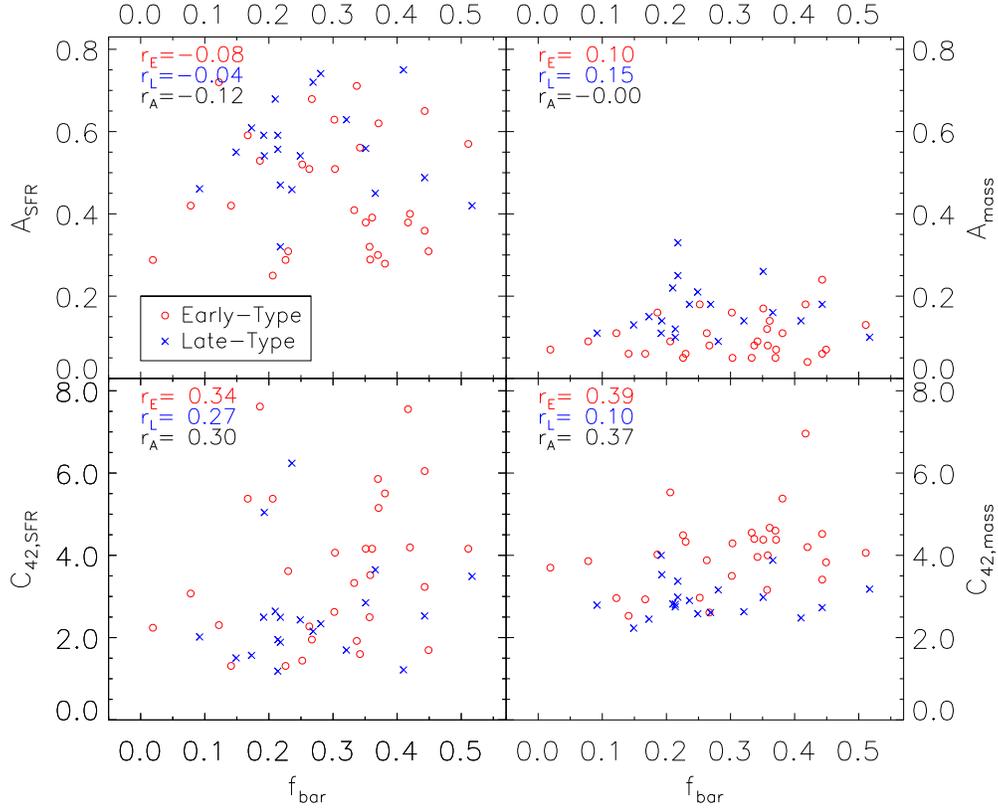} 
\caption{Relationship between bar strength and the asymmetry ({\it upper} panels) and concentration index ({\it lower} panels) in our samples, derived based on the SFR images ({\it left}) and mass images ({\it right}) in Figure~\ref{fig1}. In each panel, Spearman's rho tests are made to test the correlation between two variables, and the correlation coefficients are listed for the early-type ($r_E$), late-type ($r_L$), and whole sample ($r_A$) galaxies. The four panels use the same signals, red circles for early-type galaxies and blue crosses for late-type ones.
\label{fig6}}
\end{figure}

In Figure~\ref{fig6}, we presented the results as a function of the bar strength $f_{bar}$. The early- and late-type galaxies cover approximately the same range in $f_{bar}$, while they have different distribution in morphological parameter spaces. To determine the correlation between two variables, Spearman's rho tests are made and the correlation coefficients are listed in each panel. The right panels of Figure~\ref{fig6} show the correlations between the stellar mass distribution and bar strength. In the whole sample, there are no clear trends between the morphological parameters ($A_{mass}$ and $C_{42,mass}$) and bar strength $f_{bar}$ (Spearman's rho $r_A =$ -0.00 \& 0.30, respectively). However, early-type galaxies show higher concentration index and lower asymmetry than late-type ones, which was also found by previous studies \citep[e.g.,][]{Munoz09}. The results presented here show a downsizing signature of the formation of galactic structures \citep{Cowie96, Bundy06, Sheth08}, i.e., more concentrative galaxies mean more mature systems.

The left panels of Figure~\ref{fig6} present the variation in the distribution of star formation activity along bar strength. Similarly, no clear trend is found between them (Spearman's rho $r_A =$ -0.12 \& 0.30, respectively). In general, star formation activity in early-type galaxies are less asymmetric and more concentrated than in late-type samples, while not always well distinguished. There are tens of early-type galaxies which also have similar $A_{SFR}$ or $C_{SFR}$ as late-type ones, and most early-type galaxies with high $C_{SFR}$ are located in the high end of the bar strength. In addition, weaker bars have lower concentrical star formation activity in all type galaxies, while the star formation activity in the spirals with stronger bars would not always have high concentrical distribution, especially in early-type galaxies. This indicates that galaxies with similar bars can have differing distribution of their star formation activities.

\subsection{Global Star Formation and that in Different Structural Components}
\label{sec4.2}
\begin{figure}[!htbp]
\center
\includegraphics[angle=0,width=\textwidth]{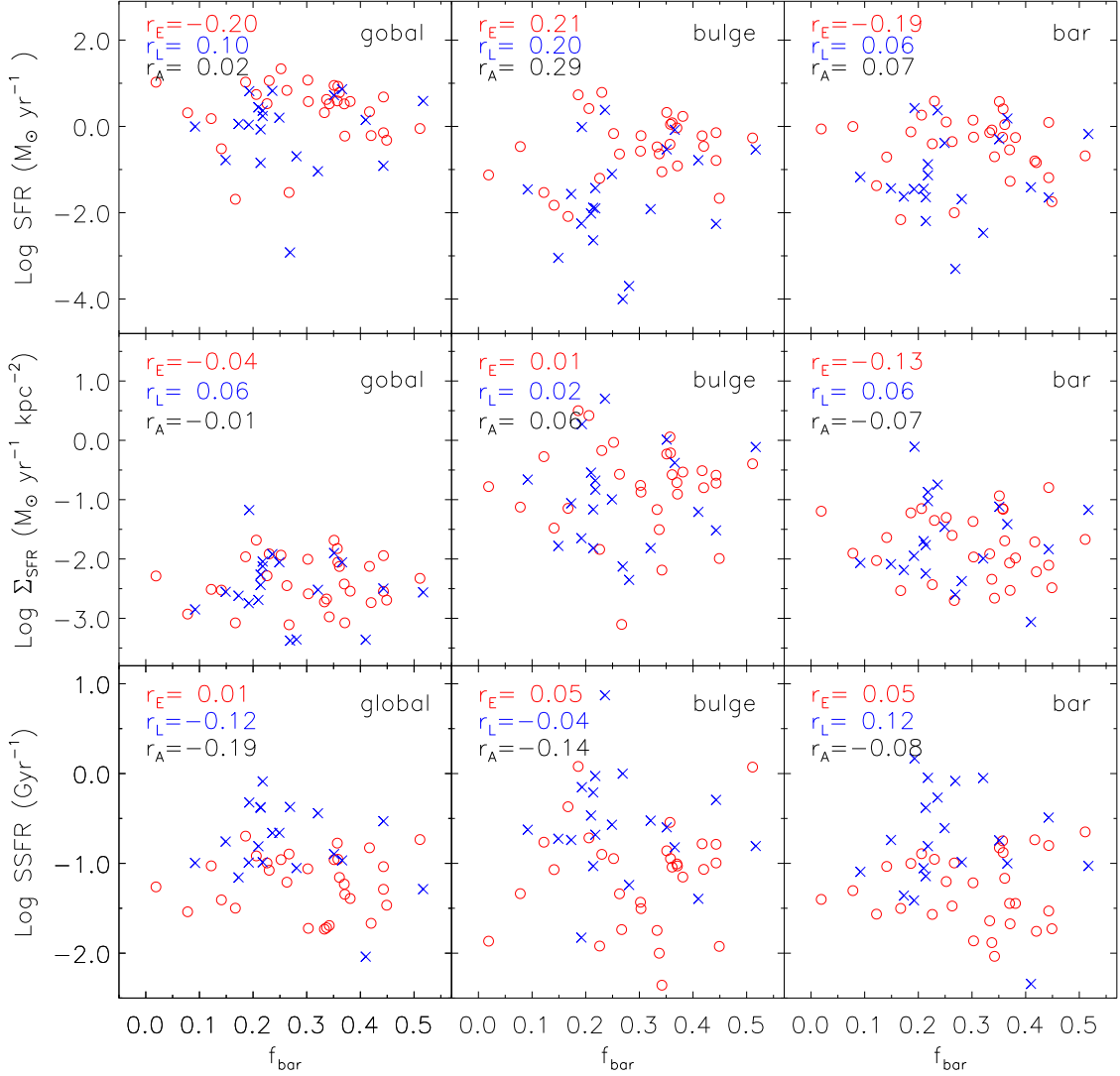}
\caption{Relationship between bar strength and star formation activity in galaxies. From {\it top } to {\it bottom}, bar strength is compared with SFR, $\Sigma_{SFR}$ and SSFR in global galaxy ({\it left}), bulge ({\it middle}) and bar ({\it right}), respectively. Spearman's rho tests are made as in Figure~\ref{fig6}. The symbols are the same as in Figure~\ref{fig2}.
\label{fig7}}
\end{figure}

In order to further explore possible relation between the star formation activity and stellar bars, we investigate the correlation between $f_{bar}$ and SFRs, SFR surface densities ($\Sigma$$_{SFR}$) and SSFRs in different galactic structural components, including bulges, bars and global galaxies. The regions of these structures are defined based on the decomposition results of the 3.6 $\mu$m images (see Sec.~\ref{sec3}). Each galaxy is divided into three structural components: bulge, bar and outer disk. In one galaxy, three structures have no overlap regions, i.e., the outer disk region is obtained by subtracting the bar and bulge field in the galaxy, and the bar region is obtained by subtracting its central bulge field. The bulge regions are defined following the decomposition results, or the galactic central kpc regions if no bulges exist in the decomposition fitting. The inclination {\it i} of galaxies can lead to underestimate of the projected area of each galactic structures, and may affect the estimate of $\Sigma$$_{SFR}$ (SFR/area). Therefore, we correcte this variation included by the inclination using the method in \citet{Martin95} (see Sec.~\ref{sec3}), and then calculate $\Sigma$$_{SFR}$. The results are listed in Table~\ref{table4}, and are plotted in Figure~\ref{fig7}.

First, the top panel of Figure~\ref{fig7} shows the SFRs as a function of the bar strength $f_{bar}$. There are no clear trends of the SFRs in each structure with the bar strength, although weak trends are found where the SFRs in bulges increase towards higher bar strength along with large scatter (Spearman's rho $r_A =$ 0.29), similar with the result in \citet{Ellison11}. Another finding is that the early- and late-type galaxies have different behaviors in the plot: (i) early-type galaxies have higher SFRs in each regions on average; and (ii) late-type ones are more various in the star formation properties, especially for intermediate-strength bars ($f_{bar}$ $\sim$ 0.2 -- 0.4). This may indicate that star formation is not a simple function of bar strength or bar-related factors, and it likely also depends on other factors \citep{Sheth02}.

Next, we compare the SFR densities and bar strength in the middle panel of Figure~\ref{fig7}. Similar to SFRs, the correlation coefficients are less than 0.2, indicating no clear trend of $\Sigma_{SFR}$ with $f_{bar}$. In addition, there is no significant difference between early- and late-type galaxies in the distribution of $\Sigma_{SFR}$. This suggests that galaxies of both types have similar $\Sigma_{SFR}$, although their SFRs are different.

Lastly, we show the correlation between SSFR and bar strength in the bottom panel of Figure~\ref{fig7}. There is a remarkable difference between the two type galaxies: most late-type samples have the SSFRs higher than 0.1 Gyr$^{-1}$, while most early-type ones have the SSFRs lower than 0.1 Gyr$^{-1}$. This trend is remarkable in the regions except bulges. The bulges have the largest scatter in the distribution of SSFRs, the highest SSFR is $\sim$ 10 Gyr$^{-1}$ in one late-type galaxy (NGC 4654), while the lowest value is nearly 4 orders of magnitude lower than it.

\section{Discussion}
A series of analysis has been provided in this work to explore the correlations between stellar bars and star formation activities in galaxies, both sub-structures and galaxies as a whole. The analysis of these correlations constrains the effect of galactic bars to star formation, reveals the secular evolution as a complex process in barred galaxies, and not simple as previously assumed. In this section, we aim to discuss the possible reasons of star formation activity varying along stellar bars. Especially, we determine how the behaviors connect to the possible episodes of bar-driven secular process. 

\subsection{The Impact of Bars on Star Formation Activity}
For barred galaxies in our sample, we had expected some evidence for enhanced star formation activity towards stronger bars. However, we find that there are no clear trends of the star formation properties along bar strength, except weak dependence on the bars for SFRs and $\Sigma$$_{SFR}$ in the bulges and global galaxies. Some previous studies also derived similar results. Using a control sample of disc galaxies, \citet{Ellison11} compared SFRs between barred and unbarred galaxies, and found that there is no correlation between bar length or bar axial ratio and the enhancement of the SFR. Some earlier studies using far infrared emission as a proxy for SFR, also did not found definite evidence that stronger bars could result in higher SFRs \citep[e.g.,][]{Pompea90, Isobe92, Roussel01}. 

But on the other hand, simulations indeed show that gas inflows driven by bars can enhance star formation activity in galactic central regions \citep[e.g.,][]{Athanassoula09}. Gas being transported to the center by bars is also found in observations \citep{Regan95, Regan99, Sheth00, Zurita08}. In addition, simulations prove that the efficiency of driving gas depends on bar strength, i.e., stronger bars are able to transport more material at a faster rate than weak bars \citep[e.g.,][]{Athanassoula92, Regan04}. 


When focusing on different types of galaxies, we find that early-type galaxies have higher SFRs, but similar $\Sigma$$_{SFR}$ and lower SSFRs compared with late-type galaxies.This may be due to the differing episodes of bar-driven gas inflow and star formation in galaxies with different types. In different episodes, there are likely differing bar stages, bulge mass, gas contents, and other physical conditions, all of which may contribute to this process \citep{OH11, Coelho11, Ellison11, Melvin14}. In late-type galaxies, stellar bars tend to be short, and bulges are usually small if exist, this is likely the early stage of bar-driven evolution. In this episode, the effect of bars are not pronounced, although there is sufficient gas in galactic disks. Within the process, the gas is reserved in the bar and galactic central regions, and then star formation activity would take place there, resulting in the growth of the stellar bar and bulge, along with the galaxy morphology changing. Therefore, we could expect the ``mature'' bars in early-type disks \citep{Debattista06, Giordano11}. 

However, the various trends of star formation properties with bar strength suggest that other process is at play in reducing bar effect: (i) massive bulges can dilute the bar nonaxisymmetric forces \citep{Das03, Laurikainen04}, thus some early-type spirals may have weaker of bars than late-type systems \citep[e.g.,][]{Laurikainen02}; (ii) the amount of gas content may be not sufficient after consumed by star formation in the previous process. Using CO observations, \citet{Sheth05} found that six early-type barred spirals have very little molecular gas detected in bars and nuclear regions due to consuming of star formation. Therefore, early-type systems may have similar or lower $\Sigma$$_{SFR}$ in their bars and central regions compared with late-type ones (Figure~\ref{fig7}). Furthermore, these probabilities are likely the reasons why various star formation properties are associated with the intermediate-strength bars. In these bars, some may be in the development from weak to strong, and others with similar strength may be weakened by the massive bulges, although they were once strong. Besides these factors, the difference between the time-scales of bar evolution and star formation can also contribute to the complex of the process \citep{Roussel01}. It is likely that the lifetime of large-scale stellar bars are long enough to allow multiple episodes of central star formation, which can interpret that why some early-type bulges have lower SSFRs (Figure~\ref{fig7}).

\subsection{The Bar-driven Secular Evolution}
\label{sec5.2}
As mentioned above, the bar-driven secular evolution is a complex process and is likely composed of a series of stages. There has been substantial progress in the scenarios of secular evolution based on simulation evidences and observational results \citep[e.g.,][]{Martin97, Combes00, Bournaud02}. 
In the evolution process, numerous episodes of bar-driven gas inflow and star formation may take place, along with the growth, destruction, and re-formation of stellar bars \citep{Combes09}. 

In one simple scenario, various stages of evolution can also be divided based on the changing properties of star formation activity and gas content in galaxies. Using properties of H$\alpha$ emission in barred galaxies, \citet{Verley07} classified them into three main groups, in which H$\alpha$ emission is mainly located in stellar bars, galactic central regions, and spiral arms, respectively. They suggested these groups as different stages of an evolution process, similar to the trend proposed by \citet{Martin97}, but more detailed. Based on the properties of circumnuclear gas and star formation, \citet{Jogee05} projected a possible picture of bar-driven dynamical evolution, where the evolution process of barred galaxies was divided into three stages (i.e., type I non-starburst, type II nonstarburst, and circumnuclear starburst). In the three stages, large amounts of gas is transformed from stellar bars to galactic central regions step by step, along with the star formation activity in bars and central regions. To complement this picture, \citet{Sheth05} supplemented a poststarburst phase, i.e., type III non-starburst, where the gas has been consumed by circumnuclear starburst, with no molecular gas in the bar region. 

Given the above evolution pictures, molecular gas is transported from galactic disks to bulges though stellar bars, resulting the rearrangement of material in the three galactic structures. Since star formation activity is correlated with the local gas density \citep{Kennicutt98}, the properties of star formation activity in bulges are likely associated with those in bars and disks. In addition, star formation activity in barred galaxies, especially in bulges, can change the galactic morphological properties (e.g., the asymmetry and concentration). To explore the correlations between them, we compared the star formation parameters of bulges, bars and disks in our sample.
\begin{figure}[!htbp]
\center
\includegraphics[angle=0,width=\textwidth]{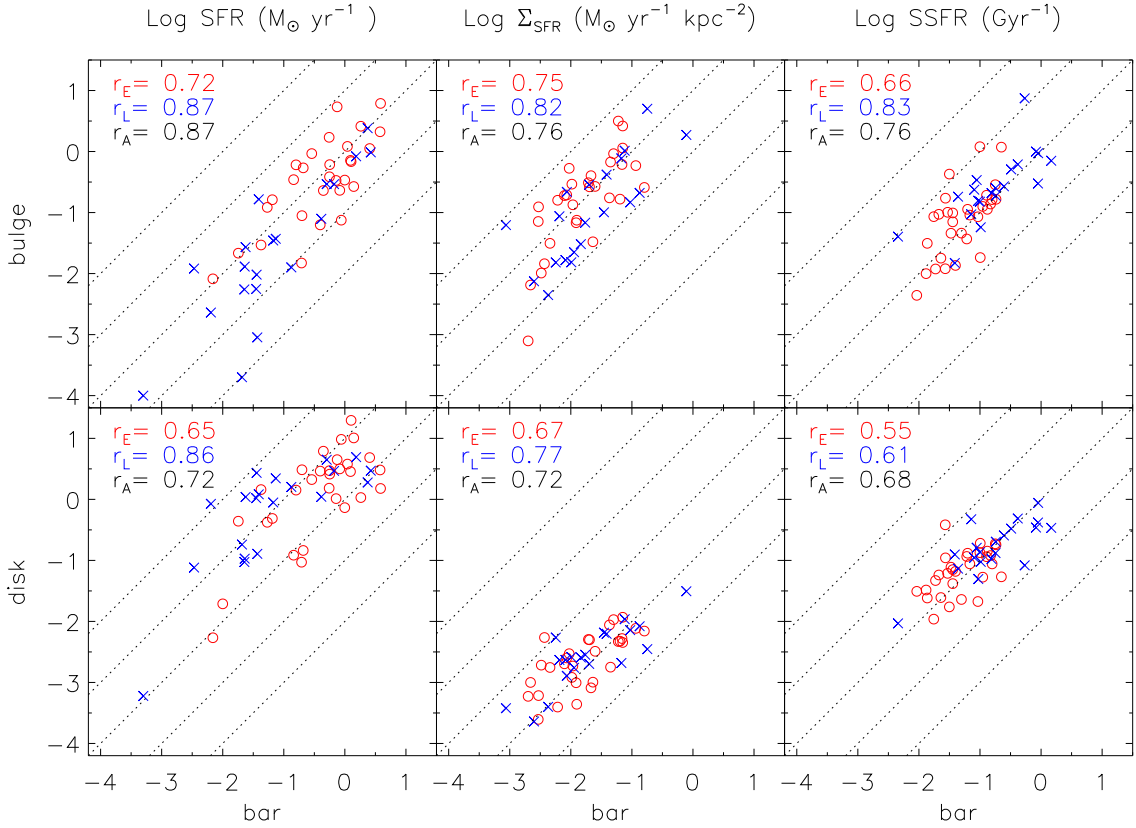}
\caption{Comparison of the SFR ({\it left}), $\Sigma$$_{SFR}$ ({\it middle}), and SSFR ({\it right}) in different structures: Bar versus bulge ({\it top}), bar versus disk ({\it bottom}). The dotted lines in each panel mark the scales between the two structures from 10$^{-2}$ to 10$^2$, top to bottom with a step of one order of magnitude. The symbols are the same as in Figure~\ref{fig2}.
\label{fig8}}
\end{figure}


Figure~\ref{fig8} illustrates the correlations between the SFRs, $\Sigma$$_{SFR}$, and SSFRs in three galactic structures, along with the ratio lines from 10$^{-2}$ to 10$^2$ plotted. We find barred galaxies with more active star formation in bars tend to have more active star formation in bulges, with similar correlation between bars and disks. It is worth noting that most SFRs in bars are 0.1 to 10 times of those in bulges, while the ratios of $\Sigma$$_{SFR}$ in bars and bulges are in the range of 0.01 to 1, one magnitude lower than the SFR ratios. In addition, the SFR densities in disks are in general 0.01 times of those in bulges. The enhanced star formation in bulges indicates the effect of bar-driven internal evolution, which can make the morphological properties vary with star formation activity in galaxies.

\begin{figure}[!htbp]
\center
\includegraphics[angle=0,width=\textwidth]{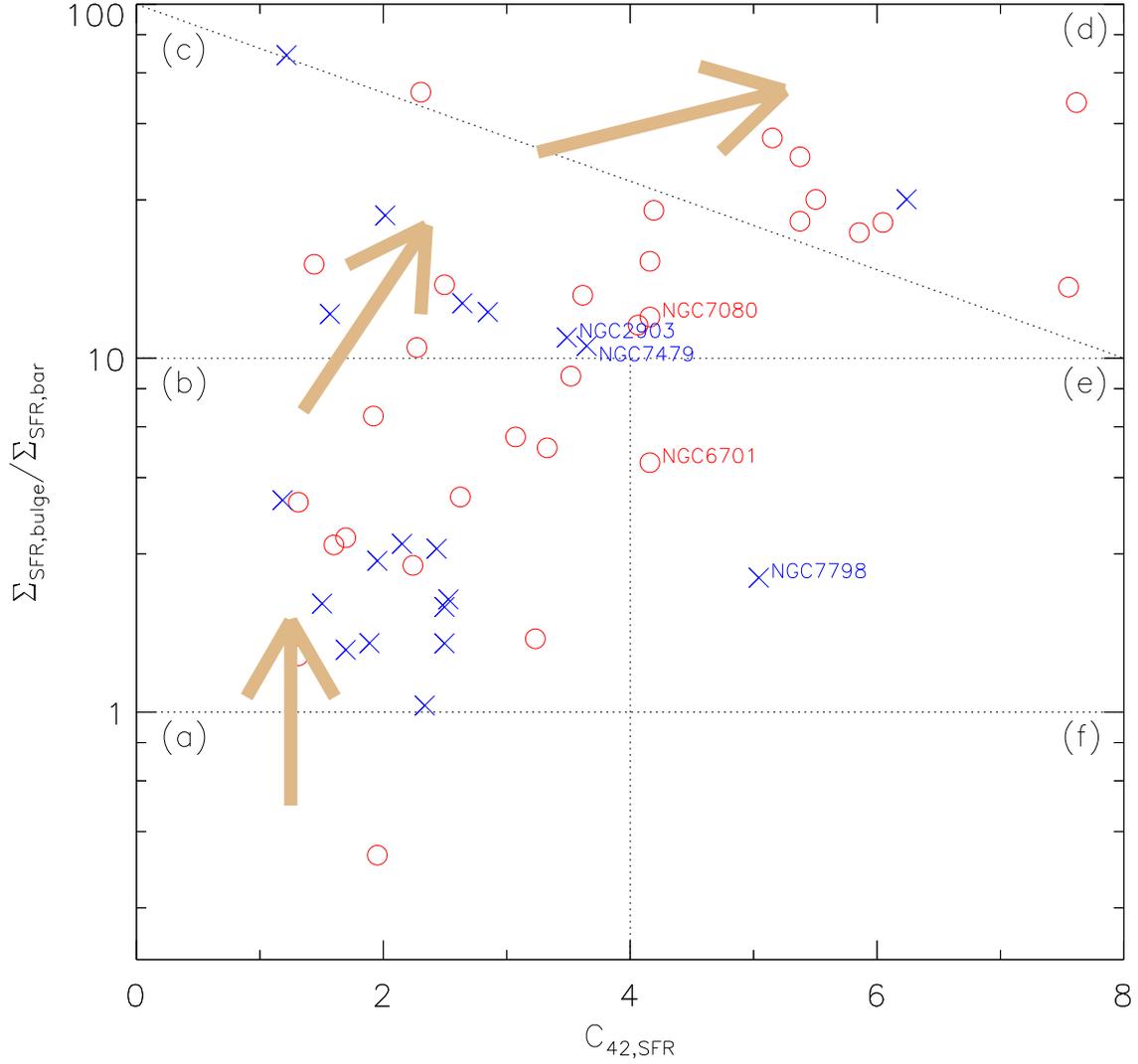}
\caption{Comparison of the ratio of bulge-to-bar $\Sigma$$_{SFR}$ with the concentration index of SFR image. This plot is used as a criterion to quantify the different stages of an bar-driven evolutionary sequence (marked with arrows from {\it a} to {\it d}). See text for detail. The arrows show an possible sequence of evolution. Five galaxies are marked specially in the plot, they are mentioned in the text or studied in our previous works \citep{Zhou11,Zhou12}. The symbols are the same as in Figure~\ref{fig2}.
\label{fig9}}
\end{figure}

Combining the above evolution scenarios and observational results, we derive a possible criterion to quantify the different stages of an evolutionary sequence in Figure~\ref{fig9}. In this figure, we use concentration index of the global SFR in galaxy, C$_{42,SFR}$, and the ratio of $\Sigma$$_{SFR}$ in bulge and in bar, $\frac{\Sigma_{SFR_{bulge}}}{\Sigma_{SFR_{bar}}}$, as the estimated parameters, and divide the parameter space to six fields (from {\it a} to {\it f}) as follows.\\
\indent {\it a}: $\frac{\Sigma_{SFR_{bulge}}}{\Sigma_{SFR_{bar}}}$ $<$ 1, C$_{42,SFR} < $ 4\\
This field is likely the early stage of bar-driven gas flow, where a considerable fraction of the star formation activity are located in galactic outer disks, and bars have more active star formation than bulges. This stage is earlier than Type I Non Starburst in \citet{Jogee05}, because most gas is still in the outer disks.\\
\indent {\it b}: $1 \leq \frac{\Sigma_{SFR_{bulge}}}{\Sigma_{SFR_{bar}}} < $10, C$_{42,SFR} < $ 4\\
This is the second stage, where the gas density in bulges are higher than that in bars along with the gas inflow. Galaxies begin to have higher active star formation densities in bulges than those in bars, while the ratio is not larger than 10, and $C_{42,SFR}$ is still lower than 4, indicating a non-negligible fraction of gas content is still in the disks. \\
\indent {\it c}: 10 $ \leq \frac{\Sigma_{SFR_{bulge}}}{\Sigma_{SFR_{bar}}} < $ 10$^{2-C_{42,SFR}/8}$\\
\indent {\it d}: $\frac{\Sigma_{SFR_{bulge}}}{\Sigma_{SFR_{bar}}} \geq $10$^{2-C_{42,SFR}/8}$\\
After most molecular gas has flowed into the galactic central regions, there are stages {\it c} and {\it d}. Field {\it c} is earlier in the process, corresponding to the later stage of Type II Non Starburst in \citet{Jogee05}, where galaxies either have low SFR concentration $C_{42,SFR}$, or have the ratio of SFR densities not high enough. Then as the intense star formation activity take place in bulges, galaxies in {\it d} may have most of star formation activity concentrated in their central regions, and also consume most of gas there, similar to the stage of Circurmnuclear Starburst in \citet{Jogee05}. \\
\indent {\it e}: 1 $\leq \frac{\Sigma_{SFR_{bulge}}}{\Sigma_{SFR_{bar}}} < $10, C$_{42,SFR} \geq $ 4\\
\indent {\it f}: $\frac{\Sigma_{SFR_{bulge}}}{\Sigma_{SFR_{bar}}} < $1, C$_{42,SFR} \geq $ 4\\
{\it e} and {\it f} have $\frac{\Sigma_{SFR_{bulge}}}{\Sigma_{SFR_{bar}}}$ lower than 10 but C$_{42,SFR}$ larger than 4. There are very few galaxies located in the two fields. Given our assumption, galaxies with low $\frac{\Sigma_{SFR_{bulge}}}{\Sigma_{SFR_{bar}}}$ are probably in the stages before circumnuclear starburst, where galaxies should have low SFR concentration. However, if there are other factors at play (such as galaxy interactions), which can promote the process of bar-driven gas inflow, large fraction of molecular gas can be concentrated in the bars and bulges of galaxies even when they are still in the early stage of bar-driven process. For example, NGC 6701 (in Field {\it e}), is apparent as isolated, but has been proved a produce of interactions based on its morphological and kinematical features \citep{Marquez96}. Similarly, NGC 7798 in Field {\it e} may be also in galaxy pairs and is affected by interactions \citep{
Chengalur93}. The resulting stages for our sample are listed in Table~\ref{table4}. 

It is noted that our scenario just presents a rough criterion with uncertainties. For example, NGC 7479 is located in Stage {\it c} in this scenario, while is suggested in the type I non-starburst evolutionary phase under the effect of a minor merger (Paper I). As discussed in Sec. 5.1, bar-driven secular evolution is a complex process, and many factors may contribute to it. To resolve this issue, more detailed observations and analysis of barred spirals are needed, along with statistical studies of large samples (e.g., using SDSS barred galaxies, Wang et al. in prep.).

\section{Summary}
We present a study of bar-driven secular evolution from a sample of 50 nearby barred galaxies with weak and strong stellar bars. In this study, we characterize the bars using image decomposition, and estimate bar strength using ellipticity-based parameter $f_{bar}$. Based on multi-wavelength photometry, we derive the properties of star formation activity in different stellar structures, and explore the correlations between these properties and stellar bars. The main results are as follows.

1. We use the image-fitting code BUDDA to perform a 2D decomposition of 3.6 $\mu$m images in bar/bulge/disc components, and obtaine the structural parameters of stellar bars. Most bars in our sample are detected less than 9 kpc, and the bars in early-type spirals are in general longer in physical and normalized lengths, and stronger in strength than those in late-type ones.

2. We calculate the rotational asymmetry A and concentration index C$_{42}$ to quantify the distributions of star formation activity and stellar mass in galaxies. These parameters are compared with bar strength, while no clear trends are found. We find that early-type galaxies have higher concentration index and lower asymmetry in both SFR and stellar mass distribution than the late-type ones. Weak bars tend to associate with low concentrical star formation activity, while strong bars appear to have large scatter in the distribution of star formation activity.

3. The correlations are explored between bar strength and SFRs, $\Sigma$$_{SFR}$ and SSFRs in bulges and bars, as well as the global galaxies, while only weak positive trends are found between star formation activity in bulges and the bar strength along with large scatter. In addition, the early- and late-type galaxies have discernible differences in their SFRs and SSFRs, but have similar $\Sigma$$_{SFR}$.

4. Comparing with previous studies, we suggest that the differing galaxy types, stellar mass distribution, and episodes of bar-driven gas inflow may contribute to the complex of the bar-driven secular evolution, which can interpret why there are no visual trends between the parameters of star formation and bar strength.

5.We find evident correlations between the star formation activity in different galactic structures, i.e., barred galaxies with intense star formation in bars tend to have active star formation in their bulges and disks, indicating the bar-driven process in galaxies.

Finally, to quantify the different stages of the process, we derived a possible criterion using the ratio of $\Sigma$$_{SFR}$ in bulge and bar, and the concentration index of the galactic global SFR as the estimated parameters. While uncertainties exist because of the contribution of other factors besides stellar bars, future works are needed to explore this issue.

\begin{acknowledgements}
We are grateful to the anonymous referee for thoughtful comments and insightful suggestions that helped to improve this paper.
We thank Yi-Nan Zhu for his generous help and assistance throughout the data observation, and M. I. Lam for advice and helpful discussions. This project is supported by Chinese National Natural Science Foundation grands No.11303038, 11433005, 11173030, 11225316, 10978014, 11073032, 11373035, 11203034, 11203031 and 11303043, by the National Basic Research Program of China (973 Program), No. 2014CB845704, 2013CB834902, and 2014CB845702, and also by the Strategic Priority Research Program ``The Emergence of Cosmological Structures'' of the Chinese Academy of Sciences, Grant No. XDB09000000.

We thank the kind staff at the Xinglong 2.16m telescope for their support during the observations. This work was partially Supported by the Open Project Program of the Key Laboratory of Optical Astronomy, National Astronomical Observatories, Chinese Academy of Sciences. This work is based on observations made with the {\it Spitzer Space Telescope}, which is operated by the Jet Propulsion Laboratory, California Institute of Technology, under NASA contract 1407. We made extensive use of the NASA/IPAC Extragalactic Database (NED) which is operated by the Jet Propulsion Laboratory, California Institute of Technology, under contract with the National Aeronautics and Space Administration.
\end{acknowledgements}


\begin{deluxetable}{ccclccrcc}
\tabletypesize{\scriptsize}
\tablecaption{Basic Parameters of the Whole Sample}
\tablewidth{0pt}
\tablehead{
\colhead{Galaxy} & \colhead{R.A.} & \colhead{Dec.} & \colhead{Morphology} & \colhead{R$_{25}$$^a$} & \colhead{{\it e}$^b$} & \colhead{P.A.$^c$} & \colhead{D$^d$} & \colhead{Source$^e$} \\
\colhead{(name)} & \colhead{(J2000.0)} & \colhead{(J2000.0)} & \colhead{(RC3)} & \colhead{(arcsec)} & \colhead{(RC3)} & \colhead{(deg)} & \colhead{(Mpc)} & \colhead{(ref)}
}
\startdata
NGC0023 & 00:09:53.40 & +25:55:26 & SBa & 62.68 & 0.35 & -30 & 58.60 & 4 \\
NGC0266 & 00:49:47.80 & +32:16:40 & SBa & 88.54 & 0.02 & 0 & 60.40 & 3 \\
NGC0337 & 00:59:50.10 & -07:34:41 & SBd & 86.52 & 0.37 & -50 & 18.30 & 1 \\
NGC1097 & 02:46:19.00 & -30:16:30 & SBb & 279.98 & 0.32 & -60 & 15.20 & 1 \\
NGC1512 & 04:03:54.30 & -43:20:56 & SBab & 267.38 & 0.37 & 90 & 11.70 & 1 \\
NGC1566 & 04:20:00.40 & -54:56:16 & SABbc & 249.53 & 0.21 & 60 & 20.60 & 1 \\
NGC2403 & 07:36:51.40 & +65:36:09 & SABcd & 656.33 & 0.44 & -53 & 2.47 & 1 \\
NGC2903 & 09:32:10.10 & +21:30:04 & SBd & 377.68 & 0.52 & 17 & 11.70 & 2 \\ 
NGC2964 & 09:42:54.20 & +31:50:50 & SABbc & 86.52 & 0.45 & -83 & 21.90 & 3 \\
NGC3049 & 09:54:49.50 & +09:16:16 & SBab & 65.63 & 0.34 & 22 & 24.70 & 1 \\
NGC3184 & 10:18:17.00 & +41:25:28 & SABcd & 222.39 & 0.07 & -45 & 11.40 & 1 \\
NGC3198 & 10:19:54.90 & +45:32:59 & SBc & 255.34 & 0.61 & 37 & 12.10 & 1 \\
NGC3274 & 10:32:17.10 & +27:40:07 & SABd & 64.14 & 0.52 & -80 & 11.50 & 2 \\
NGC3319 & 10:39:09.40 & +41:41:12 & SBcd & 184.98 & 0.45 & 37 & 13.50 & 3 \\
NGC3344 & 10:43:31.10 & +24:55:20 & SABbc & 212.38 & 0.09 & 0 & 12.30 & 2 \\
NGC3351 & 10:43:57.70 & +11:42:14 & SBb & 222.39 & 0.32 & 13 & 15.50 & 1 \\
NGC3368 & 10:46:45.70 & +11:49:12 & SABab & 227.57 & 0.31 & 5 & 17.10 & 2 \\
NGC3486 & 11:00:23.90 & +28:58:29 & SABc & 212.38 & 0.26 & 80 & 13.50 & 2 \\
NGC3521 & 11:05:48.60 & -00:02:09 & SABbc & 328.94 & 0.53 & -17 & 16.10 & 1 \\
NGC3627 & 11:20:15.00 & +12:59:30 & SABb & 273.6 & 0.54 & -7 & 14.80 & 1 \\
NGC4020 & 11:58:56.60 & +30:24:44 & SBd & 62.68 & 0.55 & 15 & 14.40 & 2 \\
NGC4051 & 12:03:09.60 & +44:31:53 & SABbc & 157.44 & 0.26 & -45 & 12.70 & 4 \\
NGC4136 & 12:09:17.70 & +29:55:39 & SABc & 119.43 & 0.07 & 0 & 12.30 & 3 \\
NGC4288 & 12:20:38.10 & +46:17:33 & SBdm & 64.14 & 0.24 & -50 & 10.00 & 2 \\
NGC4303 & 12:21:54.90 & +04:28:25 & SABbc & 193.7 & 0.11 & 0 & 26.40 & 3 \\
NGC4314 & 12:22:32.00 & +29:53:43 & SBa & 125.06 & 0.11 & -34 & 17.10 & 3 \\
NGC4394 & 12:25:55.60 & +18:12:50 & SBb & 108.92 & 0.11 & -40 & 17.00 & 4 \\
NGC4491 & 12:30:57.10 & +11:29:01 & SBa & 50.95 & 0.50 & -32 & 11.40 & 4 \\
NGC4535 & 12:34:20.30 & +08:11:52 & SABc & 212.38 & 0.29 & 0 & 31.70 & 4 \\
NGC4548 & 12:35:26.40 & +14:29:47 & SBb & 161.11 & 0.21 & -30 & 11.10 & 4 \\
NGC4569 & 12:36:49.80 & +13:09:46 & SABab & 286.5 & 0.54 & 23 & 17.00 & 1 \\
NGC4579 & 12:37:43.50 & +11:49:05 & SABb & 176.65 & 0.21 & -85 & 25.40 & 1 \\
NGC4580 & 12:37:48.40 & +05:22:07 & SABa & 62.68 & 0.22 & -15 & 18.90 & 4 \\
NGC4647 & 12:43:32.30 & +11:34:55 & SABc & 86.52 & 0.21 & -55 & 23.90 & 3 \\
NGC4654 & 12:43:56.60 & +13:07:36 & SABcd & 146.93 & 0.43 & -52 & 18.80 & 3 \\
NGC4725 & 12:50:26.60 & +25:30:03 & SABab & 321.46 & 0.29 & 35 & 20.50 & 1 \\
NGC5248 & 13:56:16.70 & +47:14:08 & SBa & 184.98 & 0.28 & -50 & 19.80 & 3 \\
NGC5377 & 13:37:32.10 & +08:53:06 & SBbc & 111.46 & 0.44 & 45 & 26.80 & 3 \\
NGC5457 & 14:03:12.60 & +54:20:57 & SABcd & 865.21 & 0.07 & 90 & 4.97 & 2 \\
NGC5713 & 14:40:11.50 & -00:17:21 & SABbc & 82.63 & 0.11 & 10 & 29.30 & 1 \\
NGC5832 & 14:57:45.70 & +71:40:56 & SBb & 111.46 & 0.41 & 45 & 6.43 & 2 \\
NGC6701 & 18:43:12.40 & +60:39:12 & SBa & 46.46 & 0.15 & -67 & 53.40 & 4 \\
NGC6946 & 20:34:52.30 & +60:09:14 & SABcd & 344.45 & 0.15 & -10 & 6.90 & 1 \\
NGC7080 & 21:30:01.90 & +26:43:04 & SBb & 54.59 & 0.05 & 90 & 62.70 & 3 \\
NGC7479 & 23:04:56.60 & +12:19:22 & SBc & 122.21 & 0.24 & 9 & 27.70 & 3 \\
NGC7552 & 23:16:10.80 & -42:35:05 & SBab & 101.65 & 0.21 & 90 & 18.80 & 1 \\
NGC7591 & 23:18:16.30 & +06:35:09 & SBbc & 58.5 & 0.57 & -35 & 63.60 & 4 \\
NGC7741 & 23:43:54.40 & +26:04:32 & SBcd & 130.95 & 0.32 & -10 & 5.53 & 3 \\
NGC7798 & 23:59:25.50 & +20:44:59 & SBc & 41.41 & 0.09 & 90 & 28.30 & 3 \\
UGC01249 & 01:47:30.60 & +27:19:52 & SBm & 207.55 & 0.55 & -30 & 0.96 & 2 \\
\enddata
\tablecomments{
$^a$ Optical radii from RC3.\\
$^b$ Ellipticity from RC3.\\
$^c$ Position angle from RC3.\\
$^d$ The luminosity distance from NED.\\
$^e$ References from where objects were selected: (1)\citet{Kennicutt03}; (2)\citet{Lee08}; (3)\citet{Pahre04}; (4)\citet{Sanders03}.
}
\label{table1}
\end{deluxetable}

\begin{deluxetable}{cccccc}
\tabletypesize{\scriptsize}
\tablecaption{XingLong 2.16m Telescope Observations}
\tablewidth{0pt}
\tablehead{\colhead{Name}& \colhead{Obs. Date} & \multicolumn{2}{c}{Exposure time (sec.)} & \colhead{H$\alpha$ filter} & \colhead{Seeing} \\
 & & H$\alpha$ & R & $\lambda$/$\Delta$$\lambda$ & (arcsec)\\
(1)&(2)&(3)&(4)&(5)&(6)
}
\startdata
NGC0023 & 2006-09-22 & 2400 & 400 & 6660/70 & 2.0\\
NGC0266 & 2008-01-10 & 3600 & 600 & 6660/70 & 2.7\\
NGC2903 & 2009-12-22 & 3600 & 600 & 6562/70 & 1.7\\
NGC2964 & 2009-01-18 & 2400 & 600 & 6610/70 & 2.0\\
NGC3319 & 2007-03-17 & 3600 & 600 & 6562/70 & 1.9\\
NGC4303 & 2007-02-24 & 1800 & 600 & 6610/70 & 2.6\\
NGC4491 & 2009-04-21 & 3600 & 600 & 6562/70 & 2.3\\
NGC4580 & 2009-04-21 & 3600 & 600 & 6610/70 & 2.3\\
NGC4647 & 2009-01-18 & 1800 & 600 & 6610/70 & 2.0\\
NGC4654 & 2008-02-14 & 3600 & 600 & 6562/70 & 2.1\\
NGC5377 & 2008-02-12 & 3000 & 600 & 6610/70 & 3.0\\
NGC5383 & 2009-04-21 & 1800 & 600 & 6610/70 & 1.8\\
NGC6701 & 2006-07-22 & 2400 & 400 & 6660/70 & 2.6\\
NGC7080 & 2006-09-23 & 2400 & 400 & 6660/70 & 1.8\\
NGC7479 & 2009-09-12 & 3000 & 600 & 6610/70 & 1.8\\
NGC7591 & 2006-09-22 & 3600 & 400 & 6660/70 & 2.4\\
NGC7798 & 2006-08-21 & 2400 & 600 & 6610/70 & 2.7\\
\enddata
\tablecomments{Col. (1): Name of the galaxy. Col. (2): The date of the observation. Cols. (3) and (4): Exposure time in each fiters in units of seconds. Col. (5): Filter used for the H$\alpha$ line, where $\lambda$/$\Delta$$\lambda$ is the central wavelength and filter width, in \AA. Col. (6): seeing in arcsec as measured on the reduced images.
}
\label{table2}
\end{deluxetable}

\begin{deluxetable}{ccccccc}
\tabletypesize{\scriptsize}
\tablecaption{Details of the archival H$\alpha$ images obtained for our sample galaxies}
\tablewidth{0pt}
\tablehead{\colhead{Name}& \colhead{Obs. Date} & \colhead{Telescope$^a$} & \multicolumn{2}{c}{Exposure time (sec.)} & \colhead{Seeing} & \colhead{Source}\\
 & & & H$\alpha$ & R & (arcsec) & \\
(1)&(2)&(3)&(4)&(5)&(6)&(7)
}
\startdata
NGC0337 & 2001-10-19 & CTIO1.5m & 1800 & 540 & 1.5 & SINGS\\
NGC1097 & 2001-10-17 & CTIO1.5m & 1800 & 480 & 1.1 & SINGS\\
NGC1512 & 2001-10-18 & CTIO1.5m & 1800 & 450 & 1.2 & SINGS\\
NGC1566 & 2001-10-21 & CTIO1.5m & 1800 & 540 & 1.3 & SINGS\\
NGC2403 & 2001-11-15 & KPNO2.1m & 1800 & 540 & 1.1 & SINGS\\
NGC3049 & 2002-04-11 & KPNO2.1m & 1800 & 420 & 1.0 & SINGS\\
NGC3184 & 2002-04-15 & KPNO2.1m & 1800 & 840 & 1.3 & SINGS\\
NGC3198 & 2002-03-09 & KPNO2.1m & 1800 & 420 & 2.3 & SINGS\\
NGC3274 & 2001-03-03 & Bok & 1000 & 200 & 1.6 & LVL\\
NGC3344 & 2001-05-27 & Bok & 1000 & 200 & 1.3 & LVL\\
NGC3351 & 2001-03-29 & KPNO2.1m & 1800 & 360 & 1.1 & SINGS\\
NGC3368 & 2003-04-29 & VATT & 1200 & 120 & 1.5 & LVL\\
NGC3486 & 2001-05-27 & Bok & 600 & 120 & 1.4 & LVL\\
NGC3521 & 2002-03-10 & KPNO2.1m & 1800 & 120 & 1.1 & SINGS\\
NGC3627 & 2002-04-14 & KPNO2.1m & 1800 & 840 & 1.1 & SINGS\\
NGC4020 & 2001-03-03 & Bok & 1000 & 200 & 1.4 & LVL\\
NGC4051 & 2000-03-31 & JKT & 3600 & 600 & 1.5 & NED\\
NGC4136 & 2001-05-16 & JKT & 3600 & 300 & 1.0 & NED\\
NGC4288 & 2001-05-01 & Bok & 1000 & 200 & 1.3 & LVL\\
NGC4314 & 1996-02-13 & INT & 1200 & 900 & 1.3 & NED\\
NGC4394 & 2001-02-17 & JKT & 3600 & 300 & 1.0 & NED\\
NGC4535 & 2000-03-06 & JKT & 3600 & 900 & 1.9 & NED\\
NGC4548 & 2000-03-31 & JKT & 3600 & 600 & 1.7 & NED\\
NGC4569 & 1988-03-25 & KPNO0.9m & 3000 & 150 & 2.0 & NED\\
NGC4579 & 2002-03-10 & KPNO2.1m & 1800 & 420 & 1.0 & SINGS\\
NGC4725 & 2002-03-11 & KPNO2.1m & 1800 & 840 & 1.6 & SINGS\\
NGC5248 & 2001-04-11 & JKT & 3600 & 1800 & 1.4 & NED\\
NGC5457 & 1995-05-20 & KPNO Schmidt & 11700 & 9545 & 2.0 & NED\\
NGC5713 & 2001-03-30 & KPNO2.1m & 1800 & 420 & 1.5 & SINGS\\
NGC5832 & 2002-03-05 & Bok & 1000 & 200 & 2.1 & LVL\\
NGC6946 & 2001-03-31 & KPNO2.1m & 1800 & 420 & 2.1 & SINGS\\
NGC7552 & 2001-10-17 & CTIO1.5m & 1800 & 450 & 1.1 & SINGS\\
NGC7741 & 2000-11-10 & JKT & 4800 & 600 & 1.5 & NED\\
UGC01249 & 2001-11-08 & Bok & 1000 & 200 & 1.7 & LVL\\
\enddata
\tablecomments{Col. (1): Name of the galaxy. Col. (2): The date of the observation. Col. (3): The telescope used. Bok means the Steward Observatory Bok 2.3m telescope on Kitt Peak; CTIO1.5m means Cerro Tololo Interamerican Observatory 1.5m telescope; INT means the 2.5-m Isaac Newton Telescope on La Palma; JKT means the 1-m Jacobus Kapteyn Telescope on La Palma; KPNO0.9m means the Kitt Peak National Observatory 0.9 m telescope; KPNO2.1m means Kitt Peak National Observatory 2.1m telescope; KPNO Schmidt means the Kitt Peak National Observatory Schmidt telescope; VATT means the Lennon 1.8m Vatican Advanced Technology Telescope on Mt. Graham, AZ. Cols. (4) and (5): Exposure time in each filters in units of seconds. (6): seeing in arcsec as measured on the reduced images. Col. (7): The sources where the data are archived. LVL: \citet{Dale09}, also part of \citet{Kennicutt08}. SINGS: \citet{Kennicutt03}. NED: the NASA Extragalactic Database (of which NGC4051, NGC4314, NGC4535, NGC4548, NGC5248, and NGC7741 are from \citet{Knapen04}, NGC4136 and NGC4394 are from \citet{James04}, NGC4569 is from \citet{Koopmann01}, and NGC5457 is from \citet{Hoopes01}).}
\label{table3}
\end{deluxetable}

\begin{deluxetable}{cccccccccccccccc}
\tabletypesize{\scriptsize}
\rotate
\tablecaption{Structural and star formation properties of the sample}
\tablewidth{0pt}
\tablehead{\colhead{Name}& \colhead{$f_{bar}$} & \colhead{$C_{42,SFR}$} & \colhead{$A_{SFR}$}& \colhead{$C_{42,mass}$} & \colhead{$A_{mass}$} & \multicolumn{3}{c}{Log SFR ($M_{\odot}~yr^{-1}$)} & \multicolumn{3}{c}{Log $\Sigma$$_{SFR}$ ($M_{\odot}~yr^{-1}~kpc^{-2}$)} & \multicolumn{3}{r}{Log SSFR ($Gyr^{-1}$)} & \colhead{stage}\\
 &&&&&&global&bulge&bar&~~~global&bulge&bar&global&bulge&bar&\\
(1)&(2)&(3)&(4)&(5)&(6)&(7)&(8)&(9)&~~~(10)&(11)&(12)&(13)&(14)&(15)&(16)
}
\startdata
NGC0023&0.23&3.62&0.31&4.33&0.06&1.06&0.79&0.58&~~~-1.91&-0.17&-1.35&-1.08&-0.90&-0.95&c\\
NGC0266&0.34&1.92&0.71&4.40&0.08&0.62&-0.64&-0.08&~~~-2.67&-1.50&-2.34&-1.72&-2.00&-1.88&b\\
NGC0337&0.25&2.43&0.54&2.58&0.21&0.20&-1.10&-0.39&~~~-2.06&-1.00&-1.46&-0.66&-0.57&-0.61&b\\
NGC1097&0.38&5.50&0.28&5.38&0.11&0.58&0.23&-0.26&~~~-2.54&-0.53&-1.98&-1.39&-1.15&-1.45&d\\
NGC1512&0.37&5.15&0.62&4.38&0.07&-0.22&-0.92&-1.27&~~~-3.08&-0.91&-2.53&-1.34&-1.03&-1.67&d\\
NGC1566&0.26&2.27&0.51&3.88&0.11&0.83&-0.64&-0.35&~~~-2.45&-0.57&-1.60&-1.21&-1.34&-1.47&c\\
NGC2403&0.22&1.89&0.47&3.37&0.25&0.24&-1.90&-0.88&~~~-2.04&-0.68&-0.87&-0.09&-0.03&-0.05&b\\
NGC2903&0.52&3.49&0.42&3.18&0.10&0.59&-0.54&-0.18&~~~-2.56&-0.11&-1.17&-1.29&-0.81&-1.03&c\\
NGC2964&0.36&2.50&0.32&3.16&0.12&0.59&-0.41&-0.25&~~~-1.82&0.06&-1.15&-0.77&-0.54&-0.75&c\\
NGC3049&0.51&4.16&0.57&4.06&0.13&-0.05&-0.27&-0.68&~~~-2.33&-0.40&-1.67&-0.74&0.07&-0.65&c\\
NGC3184&0.17&1.57&0.61&2.45&0.15&0.06&-1.57&-1.63&~~~-2.62&-1.06&-2.19&-1.16&-0.74&-1.36&c\\
NGC3198&0.09&2.02&0.46&2.79&0.11&-0.00&-1.46&-1.17&~~~-2.85&-0.66&-2.06&-1.00&-0.63&-1.09&c\\
NGC3274&0.21&1.18&0.56&2.75&0.12&-0.85&-1.89&-1.64&~~~-2.44&-1.17&-1.76&-0.38&-0.21&-0.38&b\\
NGC3319&0.28&2.34&0.74&3.16&0.09&-0.69&-3.70&-1.68&~~~-3.36&-2.35&-2.37&-1.05&-1.24&-0.99&b\\
NGC3344&0.12&2.30&0.72&2.96&0.11&0.18&-1.53&-1.37&~~~-2.51&-0.27&-2.03&-1.03&-0.76&-1.57&d\\
NGC3351&0.37&5.86&0.30&4.60&0.05&0.52&-0.03&-0.54&~~~-2.42&-0.71&-2.07&-1.23&-1.01&-1.45&d\\
NGC3368&0.33&3.33&0.41&4.55&0.05&0.32&-0.47&-0.14&~~~-2.72&-1.17&-1.91&-1.73&-1.75&-1.64&b\\
NGC3486&0.19&2.50&0.59&4.00&0.11&0.04&-2.25&-1.45&~~~-2.74&-1.65&-1.95&-0.99&-1.83&-1.41&b\\
NGC3521&0.02&2.24&0.29&3.70&0.07&1.02&-1.12&-0.06&~~~-2.29&-0.78&-1.19&-1.26&-1.87&-1.40&b\\
NGC3627&0.30&2.62&0.63&3.50&0.16&1.07&-0.57&0.15&~~~-2.00&-0.76&-1.37&-1.06&-1.43&-1.22&b\\
NGC4020&0.15&1.50&0.55&2.23&0.13&-0.78&-3.05&-1.43&~~~-2.55&-1.78&-2.09&-0.76&-0.73&-0.74&b\\
NGC4051&0.42&7.55&0.38&6.96&0.18&0.34&-0.22&-0.80&~~~-2.12&-0.51&-1.71&-0.83&-0.78&-0.74&d\\
NGC4136&0.21&1.95&0.59&2.81&0.10&-0.07&-2.64&-2.19&~~~-2.26&-1.82&-2.25&-0.37&-1.03&-1.14&b\\
NGC4288&0.32&1.70&0.63&2.63&0.14&-1.04&-1.92&-2.47&~~~-2.52&-1.82&-1.99&-0.44&-0.52&-0.05&b\\
NGC4303&0.25&1.44&0.52&2.97&0.18&1.34&-0.17&0.10&~~~-1.94&-0.03&-1.30&-0.96&-0.95&-1.20&c\\
NGC4314&0.42&4.19&0.40&4.20&0.04&-0.21&-0.46&-0.84&~~~-2.74&-0.80&-2.22&-1.67&-1.07&-1.76&c\\
NGC4394&0.44&6.05&0.65&4.52&0.06&-0.15&-0.79&-1.19&~~~-2.55&-0.72&-2.10&-1.29&-1.00&-1.53&d\\
NGC4491&0.17&5.38&0.59&2.93&0.06&-1.69&-2.09&-2.16&~~~-3.08&-1.15&-2.53&-1.50&-0.37&-1.50&d\\
NGC4535&0.41&1.22&0.75&2.48&0.14&0.15&-0.78&-1.41&~~~-3.36&-1.21&-3.06&-2.04&-1.39&-2.34&d\\
NGC4548&0.45&1.70&0.31&3.83&0.07&-0.32&-1.67&-1.74&~~~-2.69&-1.99&-2.48&-1.47&-1.92&-1.73&b\\
NGC4569&0.08&3.07&0.42&3.86&0.09&0.32&-0.47&-0.00&~~~-2.93&-1.13&-1.90&-1.54&-1.34&-1.30&b\\
NGC4579&0.30&4.06&0.51&4.29&0.05&0.58&-0.21&-0.25&~~~-2.59&-0.87&-1.97&-1.72&-1.50&-1.86&c\\
NGC4580&0.14&1.31&0.42&2.53&0.06&-0.52&-1.83&-0.71&~~~-2.52&-1.48&-1.64&-1.41&-1.07&-1.03&b\\
NGC4647&0.22&2.50&0.32&2.98&0.33&0.37&-1.43&-1.13&~~~-2.12&-0.83&-1.03&-0.99&-0.68&-0.81&b\\
NGC4654&0.24&6.24&0.46&2.90&0.18&0.82&0.38&0.37&~~~-1.92&0.70&-0.75&-0.66&0.87&-0.27&d\\
NGC4725&0.34&1.60&0.56&3.96&0.09&0.52&-1.05&-0.70&~~~-2.97&-2.19&-2.66&-1.69&-2.36&-2.04&b\\
NGC5248&0.19&7.62&0.53&4.02&0.16&1.03&0.73&-0.13&~~~-1.96&0.50&-1.22&-0.70&0.08&-1.00&d\\
NGC5377&0.23&1.31&0.29&4.49&0.05&0.53&-1.20&-0.40&~~~-2.28&-1.84&-2.43&-1.00&-1.92&-1.57&b\\
NGC5457&0.21&2.64&0.68&2.82&0.22&0.44&-2.02&-1.44&~~~-2.69&-0.54&-1.70&-0.81&-0.47&-1.05&c\\
NGC5713&0.44&3.23&0.36&3.41&0.24&0.68&-0.15&0.09&~~~-1.94&-0.59&-0.80&-1.04&-0.79&-0.80&b\\
NGC5832&0.27&1.95&0.68&2.61&0.08&-1.53&-4.78&-2.00&~~~-3.11&-3.10&-2.70&-0.90&-1.74&-1.00&a\\
NGC6701&0.35&4.16&0.38&4.38&0.17&0.95&0.32&0.58&~~~-1.68&-0.23&-0.94&-0.96&-0.86&-0.83&e\\
NGC6946&0.35&2.85&0.56&2.98&0.26&0.72&-0.53&-0.30&~~~-1.90&0.01&-1.12&-0.90&-0.60&-0.74&c\\
NGC7080&0.36&4.16&0.39&4.67&0.14&0.79&0.08&0.04&~~~-2.13&-0.58&-1.69&-1.16&-1.04&-1.17&c\\
NGC7479&0.37&3.65&0.45&3.88&0.16&0.86&-0.08&0.18&~~~-2.06&-0.38&-1.41&-0.97&-0.82&-1.00&c\\
NGC7552&0.21&5.38&0.25&5.53&0.09&0.74&0.41&0.26&~~~-1.68&0.42&-1.15&-0.92&-0.72&-0.89&d\\
NGC7591&0.36&3.52&0.29&4.00&0.08&0.93&0.05&0.41&~~~-2.05&-0.21&-1.16&-0.95&-0.95&-0.88&b\\
NGC7741&0.44&2.53&0.49&2.73&0.18&-0.91&-2.26&-1.65&~~~-2.50&-1.52&-1.84&-0.53&-0.29&-0.49&b\\
NGC7798&0.19&5.04&0.54&3.53&0.14&0.82&-0.01&0.43&~~~-1.18&0.27&-0.11&-0.32&-0.15&0.17&e\\
UGC1249&0.27&2.15&0.72&2.61&0.18&-2.92&-4.00&-3.30&~~~-3.38&-2.13&-2.60&-0.37&-0.00&-0.08&b\\
\enddata
\tablecomments{(1) Galaxy name. (2) Bar strength estimated using Eq.~\ref{eq1}. (3,4) The concentration and asymmetry index for the star formation distribution in one galaxy, described in Sec.~\ref{sec4.1}. (5,6) The concentration and asymmetry index for the stellar mass distribution in one galaxy, described in Sec.~\ref{sec4.1}. (7--15) The star formation rate, surface density of star formation rate and specific star formation rate for the galaxy as an entirety, bulge and bar, respectively, calculated as in Sec~\ref{sec4.2}. (16) The stage in bar-driven evolutionary sequence, estimated from the position in Figure~\ref{fig9} and described in Sec.~\ref{sec5.2}.  
}
\label{table4}
\end{deluxetable}

\end{document}